\documentclass[apj]{emulateapj}  
\usepackage{apjfonts}
\usepackage{mathptmx}
\usepackage{psfig}
\usepackage[usenames]{color}

\newcommand {\etal}{{et\thinspace al.} }

\def\iso#1#2{\mbox{${}^{#2}{\rm #1}$}}
\def\li#1{\hbox{$^{#1}{\rm Li}$}}
\def\be#1{\iso{Be}{#1}}
\def\b#1#2{\iso{B}{#1#2}}
\def\6Li{\mbox{$^6$Li}}
\def\7Li{\mbox{$^7$Li}}

\def\beq{\begin{equation}}
\def\eeq{\end{equation}}

\def \Sec#1{{Sec\-tion~\ref{s:#1}}}

\def \Eq#1{{Eq.~(\ref{eq:#1})}}     
\def \EQN#1{\label{eq:#1}}        
\def \Fig#1{{Fig.~\ref{f:#1}}}   
\def \Figs#1#2{{Figs.~(\ref{f:#1})-(\ref{f:#2})}} 

\shorttitle{}
\shortauthors{}

\begin{document}

\journalinfo{
\parbox{1.5in}{
UMN--TH--2612/07 \\
FTPI--MINN--07/23 \\
July 2007}
} 
\title{Cosmic Ray production of Beryllium and Boron  at high redshift}

\author{Emmanuel Rollinde\altaffilmark{1}, David Maurin\altaffilmark{1,2}, 
Elisabeth Vangioni\altaffilmark{1}, Keith A. Olive\altaffilmark{3}, Susumu Inoue\altaffilmark{4}}
\altaffiltext{1}{Institut d'Astrophysique de Paris, UMR7095 CNRS, Universit\'e Pierre et Marie Curie, 98 bis bd Arago, 75014 Paris, France, rollinde@iap.fr, vangioni@iap.fr}
\altaffiltext{2}{Laboratoire de Physique Nucl\'eaire et Hautes Energies, CNRS-IN2P3/Universit\'es Paris VI et Paris VII, 4 place Jussieu, Tour 33, 75252 Paris Cedex 05, France, dmaurin@lpnhe.in2p3.fr}
\altaffiltext{3}{William I. Fine Theoretical Physics Institute, School of Physics and Astronomy, University of Minnesota, Minneapolis, MN 55455, USA, olive@physics.umn.edu}
\altaffiltext{4}{National Astronomical Observatory of Japan, 2-21-1 Osawa, Mitaka, Tokyo 181-8588, Japan}

\begin{abstract}
Recently, new observations of \6Li\ in Pop~II stars of the galactic halo
have shown a surprisingly high abundance of this isotope, about a thousand
times higher than its predicted primordial value. In previous papers, a cosmological
model for the cosmic ray-induced production of this isotope in the IGM
has been developed to explain the  observed abundance at low metallicity.
In this paper, given this constraint on the \6Li, we calculate
the non-thermal evolution with redshift of D, Be, and B  in
the IGM. In addition to cosmological cosmic ray interactions
in the IGM, we include additional processes driven by SN explosions:
neutrino spallation and a low energy
component in the structures ejected by outflows to the IGM.
We take into account CNO CRs impinging on the intergalactic gas.
Although subdominant in the galactic disk, this process is shown
to produce the bulk of Be and B in the IGM, due to the differential
metal enrichment between structures (where CRs originate) and the IGM.
We also consider the resulting extragalactic gamma-ray background 
which we find to be well below existing data.
The computation is performed in
the framework of hierarchical structure formation considering several
star formation histories including Pop~III stars.
We find that D production is negligible and that a potentially detectable 
Be and B plateau is produced by these processes
at the time of the formation of the Galaxy ($z\sim 3$). 
\end{abstract}



\section{Introduction}
\label{s:introduction}

Big Bang Nucleosynthesis (BBN) together with Galactic Cosmic-Ray Nucleosynthesis 
(GCRN) has revealed a consistent picture for 
the origin and evolution of deuterium, helium, lithium, beryllium and boron.
This combination involves very different aspects of nucleosynthesis including
primordial, non-thermal and stellar nucleosynthesis,
all of which are correlated through cosmic and chemical evolution. 
There is one single free parameter in the standard model of BBN,
the baryon density, and that has now been determined with high precision
by WMAP \citep{Spergel} rendering BBN a parameter-free theory \citep{cfo2,cfo3}.
The derived BBN value of D/H is in agreement with that deduced 
from observations 
(D/H) = $2.84 \pm 0.26 \times 10^{-5}$ 
\citep[~and references therein]{omeara}.
This is a key success of Big Bang cosmology.

Unlike  deuterium, which is observed in high redshift quasar absorption systems,  
the LiBeB abundances are primarily 
determined from observations of the atmospheres of stars
in the halo of our Galaxy. Low
metallicity stars (Pop II stars) offer various constraints on the
early evolution of those light elements, assuming that time and 
metallicity are correlated. Until recently, the evolution of the abundances of 
\li6, \li7, $^9$Be, and $^{10,11}$B could be explained in the context of GCRN 
along with the primordial value of \li7 from BBN which forms the 
{\em Spite} plateau \citep{spites}  for stars with 
metallicity lower than about [Fe/H] =  -1.5
\citep[for a review see][]{Vangioni00}. 
The first observations of \6Li
at [Fe/H] $\simeq -2$ \citep{sln1,ht1,ht2,sln2,cetal,Nissen00}
were entirely consistent with the
predicted abundances of \6Li/\7Li $\simeq 0.05$ in standard GCRN models \citep{sfosw,Fields99,Vangioni99}. 

Unfortunately,  recent observations have led to
two distinct Li problems. First, given the baryon density inferred
from WMAP, the BBN predicted values of \7Li are 
\7Li/H = $4.27^{+1.02}_{-0.83}\times 10^{-10}$ \citep{cfo,cfo3,Cyburt04}, 
\7Li/H = $4.9^{+1.4}_{-1.2}\times10^{-10}$ \citep{Cuoco}, or
 \7Li/H = $4.15^{+0.49}_{-0.45}\times10^{-10}$ \citep{Coc}.
 These values are 
all significantly larger than most determinations
 of the lithium (\7Li +\6Li) abundance in Pop II stars which are in the range
 $1 - 2 \times 10^{-10}$
 \citep[see e.g.][]{spites,rbofn,bona}.
These values are also larger than a recent determination
by \citet{mr} based on a higher temperature scale. 
This is still an open question which will not be addressed in this paper.
Second, recent observations of \6Li
at low metallicity \citep{Asplund,Inoue} indicate
 a value of [\6Li] $= \log{\6Li/{\rm H}} +12 = 0.8$ 
which appears to be {\em independent} of
 metallicity in sharp contrast to what is expected from GCRN models,
and is more consistent with a pre-galactic origin.
While BBN does produce a primordial abundance of  \6Li, it is at
the  level of about 1000 times below these recent observations
 \citep{tsof,Vangioni99}. GCRN builds on the BBN value
 yielding a \li6 abundance which would be proportional to 
 metallicity. Thus, an additional source of \li6 is required
at low metallicity ([Fe/H] < -2) and 
different scenarios have been discussed 
which include the production of  \6Li\
during the epoch of structure formation \citep{Suzuki,Nakamura,Tatischeff}
or through the decay of relic particles during 
the epoch of the big bang nucleosynthesis
\citep[e.g.][]{kawa,Jedamzik3,grant,posp,cefos}.

In previous papers \citep[][ hereafter RVOI, RVOII]{RVOI,RVO2}, we 
investigated the possibility of high-redshift {\em Cosmological
Cosmic-Rays} (CCRs), accelerated  by  the winds of Pop III SN, and 
thus related the non-thermal production of \li6  to 
an early population of massive stars.
The star formation history was computed 
in the framework of the hierarchical structure
formation scenario, as described in \citet{Daigne05}.
RVOII showed that the production of \6Li\
by the interaction of CCRs with
the IGM provides a simple way to explain
the observed \6Li\ abundance, 
within a global hierarchical structure
formation scenario that accounts for reionization, the star formation
rate (SFR) at redshift $z \la 6$, the observed chemical abundances in damped 
Lyman alpha absorbers and in the intergalactic medium.
The additional amount of \7Li\ produced
is negligible, so that the discrepancy between the theoretical prediction 
and the observed Spite plateau is not worsened.

Here, we examine the consequences of the CCR
production of \li6 on the abundances of the related Be and B isotopes. 
The CCR spallation of p, $\alpha$ on CNO in the IGM
 as well as the reverse process
of the spallation of CNO in CCRs on H and He in the IGM are
considered in computing the IGM abundances of the LiBeB elements.
As a consistency check, we also compute the abundance of D in the IGM
and the extra-galactic $\gamma$-ray background (EGRB) produced
by the same processes \citep{ss92,Fields,Pavlidou}.
As in RVOII, we assume that the IGM abundances act as 
a prompt initial enrichment (PIE) for the halo stars when
the galaxy forms (at $z\sim 3$). The observed \6Li\ plateau 
at low metallicity is assumed to originate solely from 
production in the IGM by those cosmological CRs,
which sets the single free parameter, taken to be the
fraction of energy available for CCRs.

In addition, we include
two low energy components, as proposed by \citet{Olive,casse,Vangioni96};
the neutrino process \citep{nu1,nu2} and low-energy 
interactions of C,O and $\alpha$ with the ISM (LEC). Both 
should be operative in high redshift structures and produce LiBeB. 
These processes are, however, confined within the environment of the star
and affect mainly the ISM, while the standard CRs escape efficiently into the IGM.

In \Sec{scenario}, an overview of the basic CCR scenario is given. 
We describe the IGM production of LiBeB (\S~\ref{s:ccr-prod}),
discuss the escape efficiency of CCRs (\S~\ref{s:escape}),
and our inclusions of the primary $\nu$-process and LEC components
(\S~\ref{s:LEC}). In \Sec{results}, our results for the IGM abundances of
$^6$Li (\S~\ref{s:Lithium}), D (\S~\ref{s:D}), BeB (\S~\ref{s:beb})
and $\gamma$-rays (\S~\ref{s:gamma}) are given,
and then summarized in \Sec{discussion}.
The assumed cosmology is $\Lambda$CDM
defined by $\Omega_m=0.3$, $\Omega_\Lambda=0.7$,
$h=0.71$ and $\Omega_bh^2=0.0224$.


\section{The CCR scenario and production mechanisms}
\label{s:scenario}

Our modeling of the nucleosynthesis by CCRs is based on the 
global scenario for hierarchical structure growth and cosmic star formation,
as developed by \citet{Daigne05}. This  model is based on an 
Press \& Schechter formalism and satisfies a number of observational
constraints such as  star formation rates at $z \la 6$, reionization at high $z$, as well as 
the abundances of several trace elements in the ISM and IGM. 
Once the initial mass function (IMF) and SFR are specified, one can compute
the supernovae driven outflows which enrich the IGM. The same supernova
rate will be used here to derive the flux of CCRs.  The efficiency for the escape of 
these CCRs is discussed below in \Sec{escape}. 

The models considered in \citet{Daigne05} were all bimodal models of star formation.
Each model contained a normal mode of stars with masses between 0.1 M$_{\odot}$ and 100 M$_{\odot}$ and an IMF with a near Salpeter slope. The  SFR of the normal mode peaks at  $z \approx 3$. 
In addition to a normal mode of star formation there is a
massive component which dominates star formation at high redshift. 
Here, we consider two possibilities for the massive mode. First, as in RVOII, we consider
a model where the massive mode corresponds to stars with  masses in the range 40-100 $M_\odot$
(so-called model 1). These stars terminate as type II supernovae.
Second, we consider a model where the massive mode
corresponds to stars with  masses in the range 270-500 $M_\odot$ (so-called model 2b). 
These massive  stars are assumed to
terminate as black holes through total collapse and do not contribute to any metal enrichment
in either the ISM or IGM. 
It is, however, unclear whether these implosions are responsible for
the acceleration of cosmic rays.
Energy must get out during the collapse, but this may be entirely in the form of 
neutrinos and gravitational waves. We have not included any contribution to the 
flux of CCRs from the massive component of model 2b stars.

\subsection{CCR origin}
\label{s:ccr-prod}

CCRs are made by elements initially
present in the ISM, accelerated by the SN shock-waves,
and then expelled from the structures.
Nucleosynthesis occurs when they finally interact with the elements present in the IGM.

We compute the energy and flux of CCRs as in RVOI and RVOII. 
The fraction of the total energy
injected in CCRs is $\epsilon_{\rm CR}/100$ (where 99\%\ of the energy is 
emitted in the form of neutrinos),
with an efficiency $\epsilon_{\rm CR}$ for CR acceleration.
The total kinetic energy ${\cal E}$ per SN in CRs is,
${\cal E}_{II}=10^{51.5}\epsilon_{\rm CR}$ ergs for stars which 
leave neutron stars as remnants, i.e. stars with masses, 
 8 M$_{\odot} < m < 30$ M$_{\odot} $ and are associated with Pop II. 
For more massive stars, up to 100 M$_\odot$, 
we take the total energy of core collapse to be 
0.3 times the mass of the He core, with  
$\/M_{He}=\frac{13}{24} \cdot (m-20\,M_{\odot})$ \citep{stardeath}.
Thus, for the massive mode (associated with pop III) 
of model 1, which is dominated by 40 M$_\odot$ stars,
we have ${\cal E}_{III}=10^{52.8}\epsilon_{\rm CR}$ ergs.
In RVOII, $\epsilon_{\rm CR}$, which within reason is a free parameter for each model,
was taken to be 0.15 for model 1.

\subsubsection{CCR Spectra}
\label{s:CCRSpectra}
As in RVOI and RVOII, the source spectra are taken to be
power laws in momentum, $dQ/dp\propto p^{-\gamma}$
(\citealt{Drury,BlandfordEichler}, or \citealt{Jones94} for
a brief introduction).
The CR proton flux is normalized to the total kinetic energy
injected by the SN. 

In our scenario, we have considered a simplified
picture for CCR generation, where the source
and propagated spectra are equal.
As in the RVOI and RVOII analysis,
where the CR spectral index $\gamma=3$ was assumed
this is consistent with the limited scope of our scenario,
where the details of injection, propagation in the structure,
escape, and propagation in the IGM are advantageously described
by a single phenomenological source/propagated power law spectrum.
We will come back to and comment further on this issue in our conclusions.

Given this phenomenological approach, we choose to stay as close as possible
to existing data. Because of the lack of a sound description 
of CR spectra at all redshifts, a possible and conservative choice
for describing the CR spectra of heavy elements (CNO) is to match the observed
present-day Galactic CR spectrum. Indeed, a departure from a
pure power law is observed at low energy: this is caused by
the well-understood galactic
confinement and preferential destruction of heavier elements
compared to lighter ones (see, e.g., Fig.~1 of \citealt{Maurin}).
For simplicity, we assume a pure power law for 
$p$ and $\alpha$ with spectral index $\gamma_{\rm p}$ and $\gamma_{\rm He}$.
In addition, the spectrum for CNO is modeled as a broken power-law:
\begin{eqnarray}
  \frac{dQ_{\rm C,N,O}}{dp}\propto \left\{
   \begin{array}{ll}
\label{eq:CNO_fit}
       \displaystyle
               p^{-\gamma_{\rm p,He}} \quad {\rm if} \quad E\geq E_0\;,\vspace{4mm}\\
\displaystyle
               p^{-\gamma_{\rm C,N,O}} \quad {\rm otherwise} \;.\vspace{2mm}
   \end{array}
   \right.
\end{eqnarray}
Based on observations of the H, He and CNO spectra \citep[see Fig.13 in][]{horandel},
we set $\gamma_{\rm p}\approx\gamma_{\rm He}\equiv \gamma\sim3$, and $\gamma_{\rm C}\approx
\gamma_{\rm N} \approx \gamma_{\rm O}\sim1.5$ with
$E_0= 8$~GeV/n. Such a parametrization will prove to be crucial
to the resulting BeB production (\S~\ref{s:beb}).

In the following, all CCRs above a given energy $E_{\rm cut}=1$~MeV/n 
are assumed to efficiently escape from the structures.

\subsubsection{CCR and IGM abundances}
\label{s:ccr-igm}
The abundance pattern in the CRs is linked to the abundance in the ISM. 
In Galactic Cosmic Rays (GCRs), significant deviations
for some elements are observed
due to differing acceleration efficiencies \citep[e.g.][]{Ellison,Meyer}.
The same behavior is assumed for CCRs.
It is convenient to define the abundances 
of different species $i$=\{He, C, N, O\}
relative to H in CCRs, the ISM and the IGM:
\begin{equation}
  F_i \equiv \frac{\Phi_i }{\Phi_p}  
	   \quad \quad
	f_i \equiv \frac{n^i_{\rm ISM}}{n^H_{\rm ISM}} \quad \quad 
 {\rm and}  \quad \quad 
	\bar{f}_i \equiv \frac{n^i_{\rm IGM}}{n^H_{\rm IGM}} \;,
	\label{eq:fi_Fi}
\end{equation} 
where $\Phi_i$ is the CCR flux  for
species~$i$ and $n^i$ is the respective number density. 
The quantity $F_i$ is related to the interstellar
abundances by,
\beq
F_i = \eta_i f_i\,,
\label{eq:eta_i}
\eeq
and $\eta_i$ is
matched to the abundance pattern observed in the
Galactic Cosmic Ray fluxes 
(see Table 1 of \citealt{Meyer2}):
 $\eta_{\rm He}~=~0.7$,
$\eta_{\rm C}=8.45$, $\eta_{\rm O}=5$ and $\eta_{\rm N}=1.47$. 
Abundances in  the ISM ($f_i$) and in the IGM
($\bar{f}_i$) are
computed as in \citet{Daigne05}.  
In the following,
we will always assume $f_{\rm He}=\bar{f}_{\rm He}=0.08$.
 The abundances of metals, on the other hand, $f_{\rm C,N,O}$
and $\bar{f}_{\rm C,N,O}$, do depend on $z$.

\begin{figure}[!h]
\unitlength=1cm
\begin{picture}(9,9)
\centerline{\psfig{width=\linewidth,figure=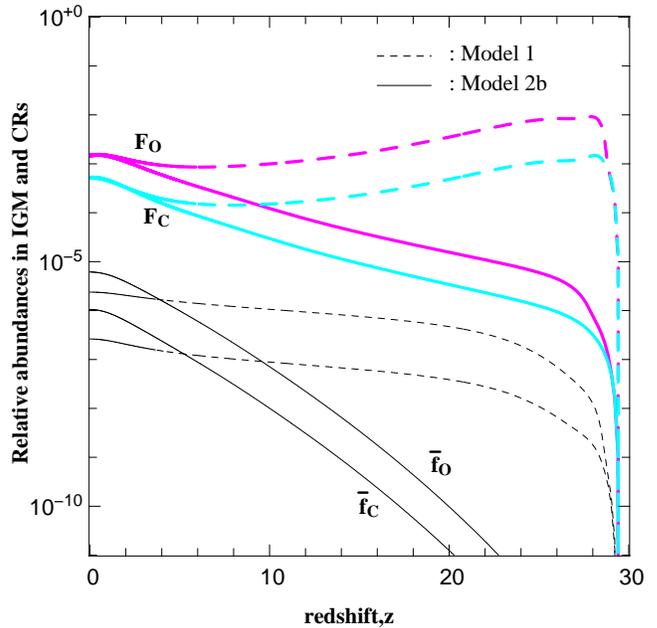}}
\end{picture}
\caption{Relative abundances of C and O in CCRs ($F_i$: colored, thick lines)
 and in the IGM ($\bar{f}_i$: black, thin lines) 
for model 2b (solid) and model 1 (dashed).
$F_i$ and  $\bar{f}_i$ are defined in Eqs.~(\ref{eq:fi_Fi}) and (\ref{eq:eta_i}).}
\label{f:CNO} 
\end{figure}

Both $F_i$ and $\bar{f}_i$ are ingredients to the IGM production
and their evolution with redshift  for C and O are displayed in  \Fig{CNO}
for the two SFR models considered. Note that the contributions of
reactions involving the CCR flux of N (not shown in the figure)
are always negligible compared to C and O. 
Note that the CR fluxes and IGM enrichment in model 2b
evolve more slowly than in model 1, due to the presumption that
the very massive Pop III stars associated with model 2b, provide neither
metal enrichment nor a source for cosmic rays.  In model 2b, both originate
from the normal mode, whereas in model 1, CRs and metal enrichment receive 
contributions from the normal mode and massive mode at higher redshift.

 \subsection{Cosmic Ray escape efficiency}
 \label{s:escape}
In RVOI and RVOII, the efficiency of secondary CR escape
from the ISM of early galactic structures (hereafter simply ``galaxies'') into the IGM
was always assumed to be high.
However, as shown in RVOII, the effects of CR heating of the IGM, 
required a certain degree of confinement of CR propagation and \li6 
production to regions that will develop into the warm-hot IGM. 
Here we present a physical justification of these assumptions.

In accordance with \citet{Daigne05},
we can evaluate the mass $\bar{M}(z)$ of a typical galaxy at redshift $z$ as the mass-weighted average
over the Press-Schechter mass function of collapsed dark matter halos $f_{PS}(M,z)$,
\beq
\bar{M}(z)=\int_{M_{\min}}^{\infty} dM M^2 f_{PS}(M,z) /  \int_{M_{\min}}^{\infty} dM M f_{PS}(M,z),
\eeq
where $M_{\min}=10^7 M_\sun$.
The average ISM density and radius of the galaxy can be estimated respectively as
\beq
\rho_{\rm ISM}(z)=(\Omega_b/\Omega_m) \Delta_c(z) \rho_c(z)
\eeq
and 
\beq
R(z)=(3 \bar{M}(z)/4 \pi \rho_c(z) \Delta_c(z))^{1/3},
\eeq
where $\rho_c(z)$ is the critical density of the universe at $z$
and $\Delta_c(z)$ is the density contrast of halos virializing at $z$
(e.g. Barkana \& Loeb 2001).

CR escape out of the ISM can be mediated either by diffusion or by advection in a galactic outflow.
The CR diffusion coefficient $\kappa$ is governed
by magnetic fields within early galaxies and is quite uncertain. 
Here we follow the plausible physical prescription of \citet{Jubelgas}  and assume
$\kappa(p,z) = 3 \times 10^{27} {\rm cm^2 s^{-1}} (p/m_p c)^{1/3}
                   (n_{\rm ISM} / {\rm cm^{-3}})^{-1/2}$,
where $p$ is the CR momentum.
Note that we have normalized to the present-day galactic value
at unit ISM number density $n_{\rm ISM}$,
and neglected the weak dependence on ISM temperature.
The timescale for diffusive escape is 
\beq
\tau_{\rm diff}(p,z)=R(z)^2/4\kappa(p,z).
\eeq
Alternatively, CRs may escape by being advected
in supernova-driven outflows of ISM gas into the IGM.
The timescale for advective escape is simply 
\beq
\tau_{\rm adv}(z)=R(z)/v_{\rm esc}(z),
\eeq
where $v_{\rm esc}(z)$ can be evaluated as in \citet{Daigne05}.

Ionization losses within the ISM may deplete LiBeB-producing, low energy CRs
before they manage to escape.
The ionization loss timescale for a CR particle of velocity $\beta c$, charge $Z$ and mass $A m_p$
in a medium of neutral H is
\begin{eqnarray}
\tau_{\rm ion}(\beta,z)  & = & (A m_p m_e c^3/ 4 \pi Z^2 e^4) \beta (\gamma-1) \\ \nonumber
                           &&    \times   [\ln(2 m_e c^2 \gamma^2 \beta^2/I_{\rm H})-\beta^2]^{-1} n_{\rm ISM}^{-1},
\end{eqnarray}
where $\gamma=(1-\beta^2)^{-1/2}$ is the particle Lorentz factor
and $I_{\rm H}=13.6$ eV is the H ionization potential (e.g. \citealt{Longair}).

\begin{figure}[!h]
\unitlength=1cm
\begin{picture}(9,9)
\centerline{\psfig{width=\linewidth,figure=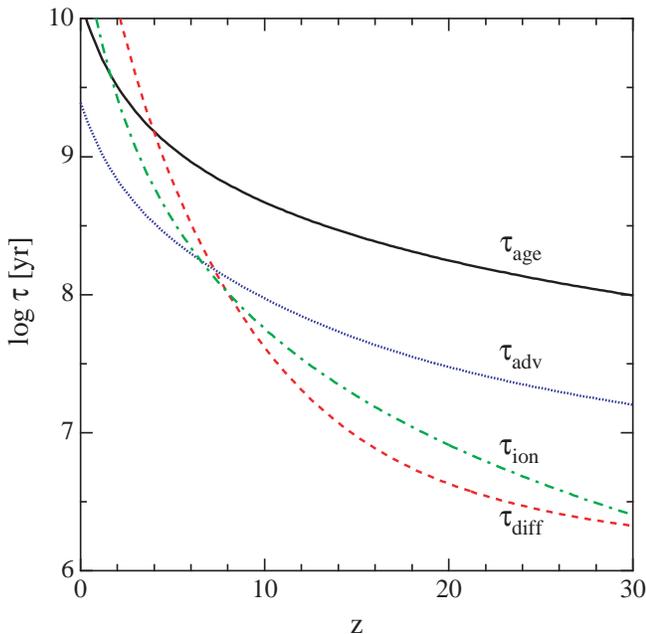}}
\end{picture}
\caption{Timescales for CR escape from typical galaxies into the IGM
due to advection $\tau_{\rm adv}$ (dotted, blue) and diffusion $\tau_{\rm diff}$ (dashed, red),
and for ionization losses within the ISM $\tau_{\rm ion}$ (dot-dashed, green),
for $\alpha$ particles of energy 30 MeV/nucleon,
compared with the age of the universe $\tau_{\rm age}$ (solid, black).
}
\label{f:tscale}
\end{figure}

In Fig.~\ref{f:tscale},
$\tau_{\rm adv}$, $\tau_{\rm diff}$ and $\tau_{\rm ion}$,
the latter two for $\alpha$ particles of kinetic energy 30 MeV/nucleon,
as functions of $z$ are compared with the age of the universe $\tau_{\rm age}$,
a rough measure of the available time at $z$.
We can see that the timescales for CR escape out of the ISM by either diffusion or advection
are always much shorter than $\tau_{\rm age}$ at all $z$,
with diffusion dominating over advection above $z \sim 7$.
This justifies the assumption of RVO2 regarding efficient CR escape into the IGM.
However, we see that in some cases ionization losses in the ISM
can become important on the escape timescales,
and such effects should be included in future, more detailed studies.

While the timescales for advection and diffusion indicate that CRs eventually
escape to the IGM, it is important to
consider the spallation timescales to determine whether there are any potential effects on the
CR spectrum.  A simple estimate of the spallation timescale, 
$ \tau_{\rm spal}= [n_{\rm ISM} \beta c \sigma_{\rm tot}]^{-1}$, shows that spallation indeed 
occurs on timescales shorter than those associated with escape.
Noting that the total destruction cross-section is roughly constant
 above a few hundred MeV and using
    $\sigma_{\rm tot}^{\rm H} \sim$ 40 mb,
   $ \sigma_{\rm tot}^{\rm He} \sim$ 80 mb, and
   $ \sigma_{\rm tot}^{\rm C} \sim $300 mb,
 one finds time scales on the order $\tau_{\rm spal} \la 10^6$ yrs.
This implies that within structures and certainly within volumes associated
with the warm-hot IGM,  we expect the effects of 
confinement to play a role and hence allows us to assume a spectrum 
with a propagated slope of $\gamma = -3$ for p's and $\alpha$'s and a 
flattened spectrum of the form given 
in Eq. (\ref{eq:CNO_fit}) for CNO. 

We note that the above estimate of $\rho_{\rm ISM}$ should become increasingly
inappropriate at lower $z$ when the total halo mass approaches Milky Way scales
and the gas within them collapses to a thin disk via efficient radiative cooling.
Then the ISM will have a much higher density within a small disk scale height, 
and much lower density for the remaining halo volume.
The CR diffusion time out of the disk should be less than $\tau_{\rm diff}$ at low $z$
as evidenced by the known CR escape time out of the current galactic disk
($\sim 10^7$ yr for GeV CRs), and that out of the halo into the IGM
should also be less by virtue of the low gas density in the halo.

\subsection{Low energy component and neutrino spallation}
\label{s:LEC}

In the past, it has been shown that, besides standard CRs, there is a need for an additional source of LiBeB in the structures for at least two reasons: standard CRs do not reproduce 
$(i)$ the meteoritic $^{11}$B/$^{10}$B isotopic ratio and $(ii)$ the linear (rather than quadratic)
proportionality between Be (and B) and Fe in the halo phase. 
These two constraints can be satisfied taking into account two additional sources of LiBeB
 \citep{Vangioni00} :
$(i)$ the neutrino spallation in the He and C shells of SN, which synthesizes $^7$Li and $^{11}$B
and $(ii)$ the break-up of low energy nuclei injected in molecular clouds which produces
all LiBeB isotopes.

Those additional processes as well as their nucleosynthesis products
take place  in the structure during SN explosion.
They are computed according to \citet{Olive,casse,Vangioni96}.
 The subsequent enrichment of the IGM
is due to the global outflow as computed by \citet{Daigne05}.

\section{Results}
\label{s:results}

For all elements $X$,
primordial abundances are assumed at the initial redshift $z=30$ of the model.
The production of $X$ by CCR interactions in the IGM 
is integrated down to $z=0$. In addition, the abundance of each
element is affected by outflows: these contributions are
shown separately in the figures.
The calculation for D, Be and B production follows
closely that of the \6Li, as given, for example, in RVOI.
The full calculation is then compared to a simplified
calculation  (Eq.~\ref{eq:efficiency_simpl} 
and values gathered in Tab.~\ref{table:xsec}; App.~\ref{app:simple}).
Note that throughout the paper,
$E$ denotes the kinetic energy per nucleon.

\subsection{Lithium} 
\label{s:Lithium}

The abundance of lithium in the IGM increases with decreasing redshift due
to the interaction of CR $\alpha$'s  with He at rest in the IGM 
(the cross section peaks at 10 MeV/n). Since the total energy in CRs is 
proportional to $\epsilon_{\rm CR}$,
the amount of lithium produced depends on both
$\gamma$ and $\epsilon_{\rm CR}$ and of course on the assumed model for star formation.
Note that lithium is also produced through CNO interactions with H and He at rest.
These are taken into account in the calculation, but the contribution to the total
abundance of Li is always subdominant in this context (\S \ref{s:CCRSpectra}).

\begin{figure}[!h]
\unitlength=1cm
\begin{picture}(9,9)
\centerline{\psfig{width=\linewidth,figure=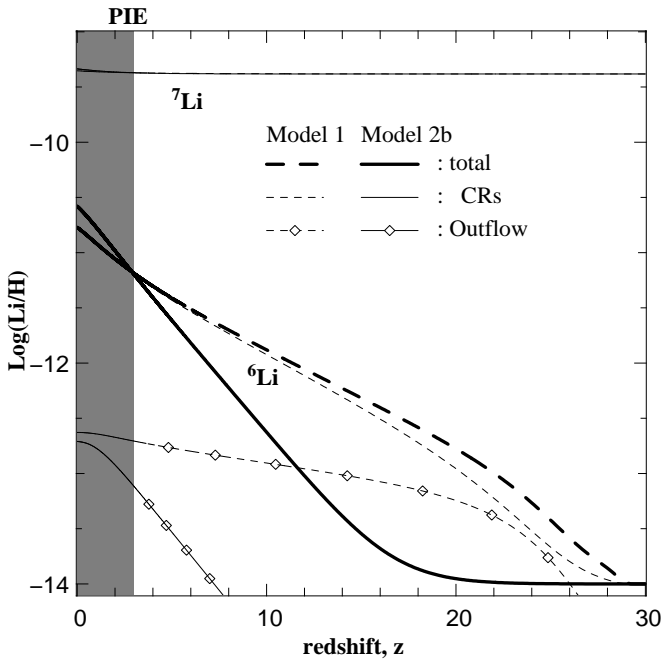}}
\end{picture}
\caption{Nucleosynthesis of both isotopes of lithium
through the interactions of secondary CCR $\alpha$ particles 
impinging on He in the IGM including the effects of outflow. 
For a given star formation history (taken here to be model 1 and model 2b 
of Daigne et al. 2006) and a 
slope of the CCR energy spectrum ($\gamma=3$), 
the acceleration efficiency of the CCRs is 
set to obtain a PIE which equals the observed 
plateau \6Li/H=$10^{-11.2}$ at $z=3$: this leads
to $\epsilon_{\rm CR} = 0.15$ for model~1 and
$\epsilon_{\rm CR} = 0.5$ for model~2b.
The individual contributions to the lithium 
abundance due to CR's and outflows from the ISM
are shown by the thin curves as indicated in the legend.}
\label{f:Li} 
\end{figure}

We have constrained
$\epsilon_{\rm CR}$ to insure a PIE of  \li6 with abundance \li6/H $= 10^{-11.2}$ at 
$z=3$. The results for model 1 and model 2b are shown in \Fig{Li}
(dashed and solid lines respectively). Model 1 was already discussed in 
RVOII and it was determined there that $\epsilon_{\rm CR} \simeq 0.15$ was necessary
to produce the requisite amount of \li6.
In model 2b, however,  PopIII stars 
terminate as black holes, with no assumed production of cosmic rays.
Thus, the lithium production is delayed and a higher efficiency 
is required, $\epsilon_{\rm CR}\simeq 0.5$. Such a high value
is consistent with those found in diffusive shock
acceleration models \citep{BV97,BE99,BV00,BV06,Blasi,Ellison05,Kang06,Ellison07},
as confirmed by recent comparison with observations of supernova remnants
\citep[e.g.][]{BV04,Warren}.

The additional lithium due to outflows from the inner regions (as discussed in \Sec{LEC})
is small  (dotted lines in \Fig{Li}) compared to the \li6 production by 
CR interactions with the IGM. 
Also shown in \Fig{Li} is the evolution of \li7 assuming the
initial primordial value of \li7/H = $4.15 \times 10^{-10}$
determined by BBN at the WMAP value for the baryon density
\citep{Coc}: as already emphasized in RVOI, the additional CCR production of \li7
is negligible compared to that produced by BBN.

\subsection{Deuterium}
\label{s:D}

There are two main production reactions for deuterium,
namely pH and pHe. For the simplified calculation, 
assuming  $\gamma=3$ and the two extreme cases
for $I(\gamma,p_1,p_2)$, as given in Eq.~(\ref{eq:Iapprox}),
we obtain
\begin{equation}
  \left[\frac{{\rm D}_{\rm pH}}{\6Li}\right] \approx 5 
	  \quad {\rm and}  \quad 
	\left[\frac{{\rm D}_{\rm pHe}}{\6Li}\right] \approx 50\,.
\end{equation}
This makes a total contribution of about D/\li6 $\sim 100$ coming from
pH, pHe and the numerically similar reverse process $\alpha$p,
 independent of the redshift. 
This is consistent with the constant value 
of 70 derived in the exact calculation.
Thus, the additional production of deuterium
by CCRs at $z=3$ is about $3.1\times 10^{-10}$  and 
is therefore well below the observed abundance D/H$\,\sim\, 2 \times 10^{-5}$
\citep{omeara}.

\subsection{BeB}
\label{s:beb}

\begin{figure}[!h]
\unitlength=1cm
\begin{picture}(9,10)
\centerline{\psfig{width=\linewidth,figure=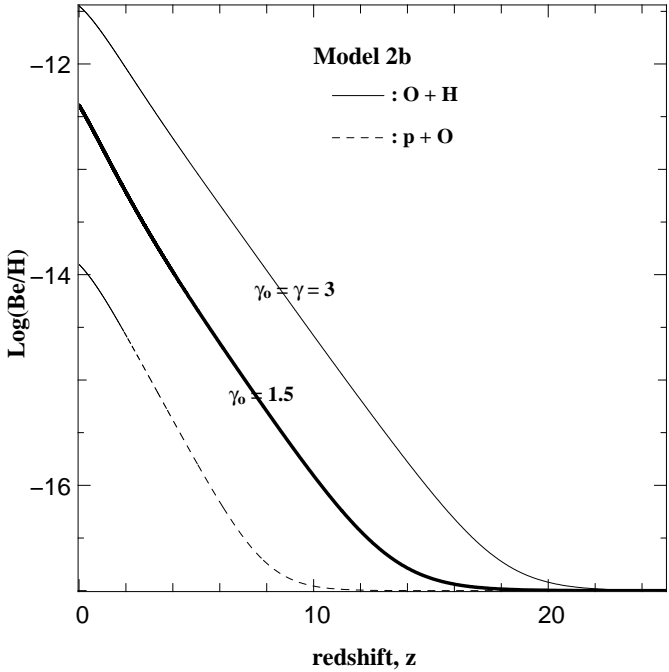}}
\end{picture}
\caption{Beryllium production by the interaction of
CCRs and the IGM for model 2b. The direct reaction p+O (dashed line)
is negligible compared to the reverse reaction O+H (solid thick line).
The upper thin solid curve displays the
\be9\ production as a function of redshift
assuming an O spectrum with $\gamma_{\rm O}$ = 3 as for p and He at low energy
The thin  dashed curve assumes  $\gamma_{\rm p, He}$ = 3. The thick solid
curve is based on
Eq.~(\ref{eq:CNO_fit}) and assumes   $\gamma_{\rm O}$ = 1.5 at low energy. 
}
\label{f:allBe} 
\end{figure}

Beryllium and boron are produced via 
spallation interaction between protons or $\alpha$'s in  CRs and
CNO elements in the IGM. The reverse reactions, which correspond to
CNO CRs spalling on H and He of the IGM,
is in fact dominant. 
This can be understood
as the metallicity in the ISM (where CRs originate)
is always larger than that in the IGM.
In contrast to the galactic disc, the heavy nuclei
component of CCRs becomes important due 
to the metal deficiency of the IGM.  Qualitatively,
the ratio of the forward to reverse processes is given by
\begin{equation}
  \left[\frac{({\rm BeB})_{\rm pO}}{({\rm BeB})_{\rm OH}}\right]= \frac{\bar{f}_{\rm O}}{F_{\rm O}}
	= \frac{[{\rm O/H}]_{\rm IGM}}{\eta_{\rm O}  [{\rm O/H}]_{\rm ISM}}\;,
\EQN{BeBsimple}
\end{equation}
where $\bar{f}$, $F$ and $\eta$ are defined in \Eq{fi_Fi} and \Eq{eta_i}.
From the calculated abundances of \citet{Daigne05}, the
metallicity ratio between the IGM and CRs is $\lesssim 10^{-2}$
(see eg. \Fig{CNO}),
so that, the forward process is $\sim 10^{-3}$ smaller than the
reverse. For the reverse process,
Eq.~(\ref{eq:efficiency_simpl})
does not apply in principle, since $F_i$ no
longer factors out of the $z$ integrand. However,
in the CRs, $F_C\lesssim 10^{-4}$ and 
$F_O\lesssim 10^{-3}$ 
(see \Fig{CNO}).
An upper limit to the production of $^9$Be, 
would then be, for $\gamma_{\rm O}=3$, 
${^9{\rm Be}_{\rm OH}}/{\6Li} \approx 0.3$.
This compares well with the full calculation (upper 
solid line in \Fig{allBe}).
The leading contribution of the reverse process
is confirmed in \Fig{allBe} for the full calculation in model 2b (upper solid
line compared to dashed line) ; the same is also true
for model 1.
 The same reasoning holds for reactions involving C, N.
The dominant channel is actually 
O+H. Nevertheless, all nuclei are accounted for in the full calculation
below.
As one can see in \Fig{allBe}, adopting the spectral break, i.e. $\gamma_{\rm O}=1.5$
 as described in \Sec{CCRSpectra}, decreases the contribution of the
 reverse process and as a result lowers the Be abundance as shown by the thick
 solid curve compared to the upper thin curve with $\gamma_{\rm O}=3$.

\begin{figure}[!h]
\unitlength=1cm
\begin{picture}(9,9)
\centerline{\psfig{width=\linewidth,figure=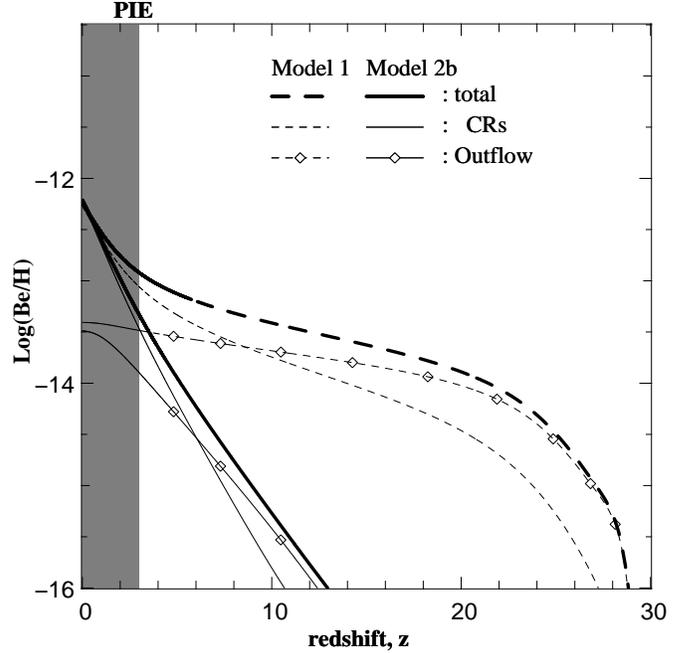}}
\end{picture}
\caption{Nucleosynthesis of beryllium and boron through the interaction 
of secondary CR p's, $\alpha$'s and (CNO) impinging on CNO (H and He) in the IGM
including the effects of outflow. 
The evolution of the abundance of $^9$Be and $^{10,11}$B
is displayed as a function of redshift 
for models 1 and 2b.
The individual contributions to the beryllium 
abundance due to CR's and outflows from the ISM
are shown by the thin curves as indicated in the legend.}
\label{f:Be} 
\end{figure}

\begin{figure}[!h]
\unitlength=1cm
\begin{picture}(9,9)
\centerline{\psfig{width=\linewidth,figure=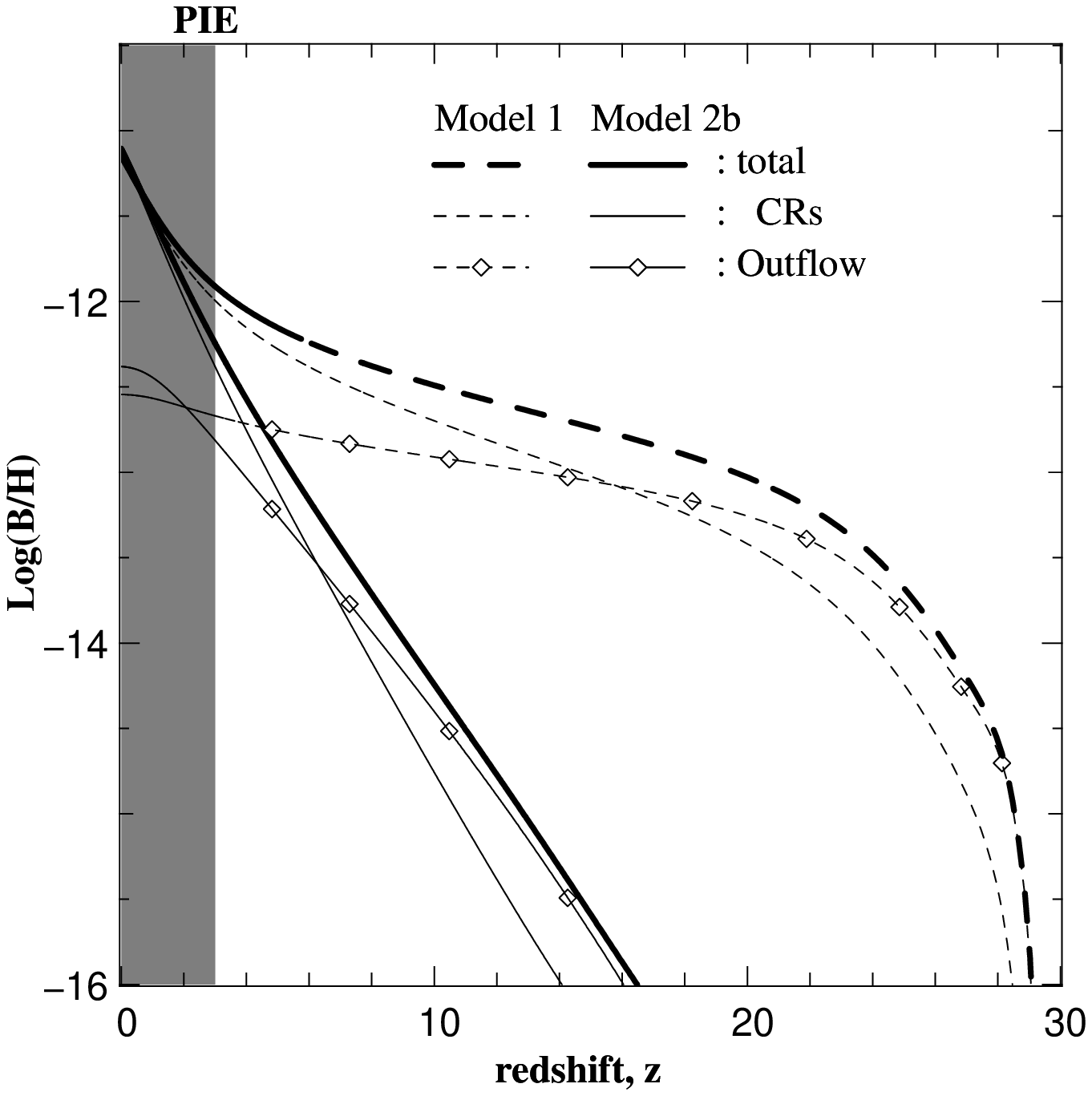}}
\end{picture}
\caption{As in \protect\Fig{Be}, for boron. }
\label{f:B} 
\end{figure}

The results from the full calculation of the BeB abundances for both models 1 and 2b
are given in \Figs{Be}{B}.  
Note once again that BeB production in model 2b is delayed due to the lack of metal  
enrichment in the ISM from the massive mode in this model.
 At $z=3$, we find \be9/H$= 10^{-12.9}$ in model 1 (upper panel)
 and \be9/H$= 10^{-13.3}$ in model 2b (lower panel).
 This is  compatible with 
 the observed abundance at the lowest metallicity, [Fe/H]$ = -3.3$,
$^9$Be/H$\sim\, 10^{-13}$  in G64-12 \citep{primasa}. 
For boron, we find
B/H $= 10^{-11.9}$ in model 1  and
 and B/H $= 10^{-12.25}$ in model 2b. These values are somewhat below
 the observed boron abundance at low metallicity
B/H$\sim\ 10^{-11.3}$ at [Fe/H]$ = -3.0$ in BD -13 3442 \citep{primasb}
though B/H is as low as $10^{-12}$ at [Fe/H]$ = -2.9$ in BD -23 3130 \citep{gletal}.
Note however, that unlike beryllium abundances, the boron abundances in particular are very sensitive
to the assumed effective temperature.  The abundances quoted above
were determined using temperatures based on the IRFM \citep{alonso}.
The boron abundance in BD -13 3442 would be $10^{-10.7}$ \citep{fov}  had we adopted 
the temperature derived by \citet{mr}. Of course Li also scales with temperature
and at higher effective temperatures, the level of the \li6 plateau would
be raised forcing one to large CR efficiencies.

Our results can be cast in terms of abundance ratios of \be9/\li6
and B/\li6
\begin{equation}
	\left[\frac{^9{\rm Be}}{\6Li}\right] \approx 0.020 
	 \qquad \left[\frac{\rm B}{\6Li}\right] \approx 0.20 \;,
	\end{equation}
for model 1 and 
\begin{equation}
\left[\frac{^9{\rm Be}}{\6Li}\right] \approx 0.008 \qquad
	\left[\frac{\rm B}{\6Li}\right] \approx 0.09 \;.
\end{equation}
for model 2b.
These predicted ratios are independent of the efficiency $\epsilon_{\rm CR}$
and can be compared to the few examples
where observational determinations of \li6 and Be exist in the same star.  
Restricting our attention to only those stars with [Fe/H] < - 2.6, we have Be 
abundance measurements in two stars, G64-12 \citep{primasa} and LP 815-43 \citep{primasb}
for which there are reliable \li6/\li7 determinations \citep{Asplund,Inoue}.
When corrected for stellar temperature differences, we find 
\be9/\li6 = $0.005 \pm 0.003$ for G64-12 and = $0.011 \pm 0.007$ for LP 815-43.
Unfortunately there are no very low metallicity stars with both 
\li6 and B abundance measurements.

The predicted PIE at $z=3$ for both elements is 
of the same order as
the abundances observed at the lowest metallicity.
Thus we expect a plateau in both Be and B
at low metallicity similar to that found for \li6.
Indeed, when the Be abundance in G64-12 was first reported
\citep{primasa}, the authors claimed evidence for a
flattening of the observed Be evolutionary trend at low metallicity.
It will be interesting if future determinations of Be at low metallicity
confirm the existence of a Be plateau.

We have also computed the production of BeB through LEC 
nucleosynthesis and neutrino spallation
via outflow to the IGM. In all cases, the abundance produced by
these processes is less at small redshift,
 though the abundance of \b11 produced by the 
$\nu$-process is not  negligible at higher redshift.
 These results are displayed by the thin lines in \Figs{Be}{B}.

\subsection{Pionic production of gamma-rays}
\label{s:gamma}

\begin{figure}[!h]
\unitlength=1cm
\begin{picture}(9,9)
\centerline{\psfig{width=\linewidth,figure=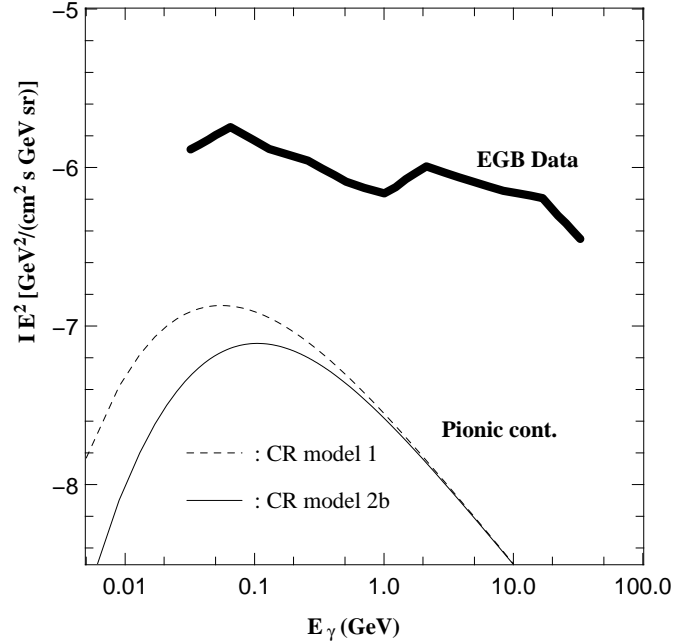}}
\end{picture}
\caption{Energy spectrum (E$^2$I versus energy  in GeV) of 
photons produced by pion decay in the IGM. The models are well
below the observed EGBR data \citep{strong}.}
\label{f:Ig} 
\end{figure}

The interaction of 
the CCRs with the IGM also produces gamma-rays.
Our calculation closely follows that given in \citet{Pavlidou}.
Assuming isotropic production,
the extragalactic differential flux of $\gamma$-rays at Earth
in the flat $\Lambda$CDM cosmology is
\begin{equation}
\frac{dF_\gamma}{dE} = \frac{c}{4\pi H_0} \int_0^\infty dz \;
\frac{q_\gamma(z,E'=(1+z)E))}{(1+z)^3\,\sqrt{\Omega_\Lambda+(1+z)^3\Omega_{\rm M}}}\;.
\end{equation}
The $\gamma$-ray source function $q_\gamma(z,E')$ is calculated
using the simplified analytic formula of Pfrommer \& Ensslin (2004):
\begin{eqnarray}
q_\gamma(z,E') &\approx& \sigma_{\rm pp} \phi_{\rm p}(E') n_{\rm IGM}(z)
\zeta^{2-\alpha} \nonumber \\
&\times& 
\frac{4m_{\pi^0}}{3\alpha_\gamma}
\times\left[ \left(\frac{2E'_\gamma}{m_{\pi^0}}\right)^\delta_\gamma
+ \left(\frac{2E'_\gamma}{m_{\pi^0}}\right)^{-\delta_\gamma}   \right]^{-\alpha_\gamma/\delta_\gamma}
\end{eqnarray}
with $\zeta=2$, $\delta_\gamma=0.14 \alpha_\gamma^{-1.6}+0.44$
and $\sigma_{\rm pp}=32\times(0.96 + e^{4.4-2.4\alpha_\gamma})$~mb.
The same input fluxes and IGM densities are taken as for the 
D and LiBeB calculations.
\Fig{Ig} shows that the total $\gamma$-ray flux
produced by CCRs  is negligible compared to
the observed EGRB. 


\section{Discussion and conclusions}
\label{s:discussion}

Our initial motivation to study CCRs at high redshift
was to find a physical origin for the anomalously high abundance of \6Li\
observed in metal poor halo stars. A cosmic history for structure formation
that reproduces standard observations \citep{Daigne05}
provides enough energy in SN to produce a \6Li\ plateau at the
level of $\log[\6Li/{\rm H}]\approx -11.2$ via $\alpha+\alpha$
interactions of CCRs with the IGM. Note that the \6Li\ produced
by outflows (via neutrino spallation and LEC) is negligible.

In this paper, we considered two models of star formation histories differing
by their Pop~III stars. The massive component of model~1 ($40-100 M_\odot$)
injects a lot of energy in CCRs so that the necessary fraction $\epsilon_{\rm CR}$
is low ($\sim 0.15$). On the other hand, the stars associated with the
massive component of model~2b ($260-500 M_\odot$) are assumed to collapse
into black holes without injecting, a priori, any energy in CCRs.
Consequently, the fraction $\epsilon_{\rm CR}$ associated with SN 
in the normal mode must be higher ($\sim 0.5$).
This figure would be reduced if super massive stars produce
CRs.

We confirm, as in RVOII, that the \7Li\ primordial abundance 
always dominates any additional production by CCRs. The same conclusion holds
for deuterium. Similarly, the observed EGBR is much larger than the intensity
of photons produced by the decay of pions produced by CCR proton collisions. 

In the same cosmological context, we explored the production
of other light elements (BeB). To that end, it is necessary to 
track the abundances of metals (CNO)
in the ISM and IGM. We have shown
that the reverse process (i.e. CNO CCRs on H and He IGM gas) is 
the dominant channel to synthesize BeB. This is in contrast
to the production of BeB in the galactic disk, where the reverse
process contributes roughly 20\%
\citep{MAR}. This is easily understood
as the metal enrichment in CCRs is inherited from structure
abundances, which are far higher than those in the IGM. Note that we also
checked that the neutrino spallation and LEC processes
are negligible in the IGM BeB budget at the time of
the galactic formation, i.e. $z\sim 3$.
In all models considered, we have shown that BeB synthesized at $z\sim 3$
is at the level of the observed abundances in the lowest metallicity
stars. This is the first theoretical indication of a  {\em plateau}
for these elements which does not resort to exotic models of BBN.

Note that these results have been obtained with several
assumptions about the CCR spectra. To some extent, our modeling,
be it for light elements (pHe) or metals (CNO),
is limited. Indeed, two fundamental ingredients for the calculation are
i) the low energy form of the source spectra for protons
and ii) the propagated fluxes to plug in the IGM. The first item is crucial
for determining the acceleration efficiency $\epsilon_{\rm CR}$. For example,
changing $\gamma\approx 3$ to $\gamma\approx 2$---hence taking a more
conventional spectral index for the sources---would lead to 
an unphysical value of $\epsilon_{\rm CR}>1$.
However, we expect that propagation effects on scales of order the
warm-hot IGM would lead to a steeper spectrum (i.e. $\gamma = 3$
for p's and $\alpha$'s) and a flatting of the CNO spectrum 
of the form we have assumed here.
The second item involves
several issues that are intimately connected: 
the details of CR escape and confinement between the
structures, the warm-hot IGM and cooler IGM is related to whether or not
the propagated spectrum displays a spectral index close to
the standard source index ($\gamma\approx 2$) or closer to a
diffused spectrum ($\gamma\approx 3$) and, more importantly,
how this evolves with $z$. 
Without some degree of confinement on the scale of the warm-hot IGM, 
we would be forced to take the same
spectrum for p, He and CNO nuclei---as would be more natural
for a 100\% escape to the IGM, and this would lead to an overproduction
of $^9$Be. Our simple estimate of the spallation timescale within
structures indicates that confinement is indeed playing a role
and changes to the CR source spectrum will occur.

All of these elements clearly call for a more coherent and refined
calculation. As just outlined, one issue concerns the description
of CR propagation in realistic structures, evolving with redshift,
which also allows for differentiated production in
situ and outside structures.
Another important issue is the possibility for heterogeneity
in metallicities at a given time (e.g. \citealt{Salvadori06}),
that cannot be handled in the homogeneous paradigm
developed in \citet{Daigne05}.  This question
could be addressed when \6Li\ and $^9$Be are observed simultaneously.
Finally, we note that B can be observed directly in high redshift objects,
as achieved for a damped Ly$\alpha$  system at $z=2.6$ \citep{prochaska}.
Observations at even higher $z$ may be possible through absorption
lines in gamma-rays bursts (J. Prochaska, private communication),
which should provide very valuable constraints on early LiBeB
production scenarios.


\acknowledgements
We warmly thank Don Ellison, Torsten Ensslin,  and Tom Jones for their help.
We thank E. Thi\'ebaut, and D. Munro for freely 
distributing his Yorick programming language (available at  {\em \tt
ftp://ftp-icf.llnl.gov:/pub/Yorick}), which we used to implement
part of our analysis. The work of EV and KO has been supported by 
the collaboration INSU - CNRS France/USA. 
The work of K.A.O. was partially supported by DOE grant
DE-FG02-94ER-40823. 
\vskip 1.cm
%


\appendix
\section{Simplified calculation for D-Be-B}
	\label{app:simple}
The calculation for D, Be and B production follows
closely that of the \6Li, as given, for example, in RVOI.
In this appendix, we estimate the relative efficiency of various channels
as well as the relative efficiency of the D, Be and B production
with respect to the \6Li\ production. In some cases, the latter is independent
of $z$. Below, $E$ denotes the kinetic energy per nucleon.

The abundance of the element $X$ relative to $H$ is given by
\begin{equation}
\left[\frac{X}{H}\right]_{z_0} = \sum_{i=CR} \; \sum_{j=IGM}
 \int_{z_0}^\infty \int_{E_{\rm cut}}^\infty 
   \bar{f}_j \; \sigma_{ij}^X(E_{\rm p}) \; F_i \;
	 \Phi_{\rm p}(E_{\rm p},z) \; dE_{\rm p} \left| \frac{dt}{dz}\right| dz\;,
\end{equation}
where $\Phi_{\rm p}(E_{\rm p},z)$ is the accumulated CR proton flux in the IGM
(in proper units), $\sigma_{ij}^X(E_{\rm p})$
is the production cross section of $X$ in the reaction $(i+j)$
and $f_i$ and $F_i$ are defined in Eqs.~(\ref{eq:fi_Fi}) and (\ref{eq:eta_i}).
Note that \6Li is only produced by CR $\alpha$'s with energy
four times the final lithium energy. As a consequence, the above formula
also applies to [$^6$Li/H], but an extra factor
$1/4$ must be added.
Two simplifications are made:
\begin{itemize}
  \item All cross sections $\sigma_{ij}^X(E)$ are approximated
	  as a constant in the range $E_1- E_2$ and zero elsewhere, i.e.
	  \begin{equation}
		  \sigma_{ij}^X(E) = \sigma_{ij}^X \times \Theta(E-E_1) \Theta(E_2-E)\;.
  	\end{equation}
		The values of the cross sections for the main processes are given in Tab.~\ref{table:xsec}.
	\item $\Phi_{\rm p}(E_{\rm p},z)$ is assumed to keep the same power law dependence 
	  $\gamma$ as the source spectrum at each $z$, so that it can be rewritten as
	  \begin{equation}
		  \Phi_{\rm p}(E_{\rm p},z) = N_0(z) \Phi_{\rm p}(E_{\rm p}) \;.
  	\end{equation}
\end{itemize}
\begin{table}
\centering
\begin{tabular}{l c c c}
\hline\hline
 $i+j \rightarrow X$   & $\sigma_{ij}^X$ (mb)  & $E_1$ (GeV/n) & $E_2$ (GeV/n)    \\\hline
 $\alpha \!\/$ + He $\rightarrow \6Li$  & 20     & 0.01  & 0.02 \\
 p + H~~~$\rightarrow$~~~D  & 1     & 0.4  & 0.8 \\
 p + He $\rightarrow$~~~D  & 12     & 0.05  & $\infty$ \\
 p + C~~~$\rightarrow \,^9$Be  & 6     & 1.0  & $\infty$ \\
 p + O~~~$\rightarrow \,^9$Be  & 5     & 0.05  & $\infty$ \\
 p + C~~~$\rightarrow $~~~B  & 90     & 0.015  & $\infty$ \\
 p + O~~~$\rightarrow $~~~B  & 50     & 0.04  & $\infty$ \\
\hline
\end{tabular}
\caption{Simplified description of cross sections used in the approximate calculation}
\label{table:xsec}
\end{table}

If $j={\rm H,~He}$, $\bar{f}_j$ and $F_i$ are constants that
factor out of the $z$ integrand: 
\begin{equation}
\left[\frac{X}{H}\right]_{z_0} = \sum_{i=CR} \; \sum_{j=IGM} 
	\bar{f}_j \; F_i \; \sigma_{ij}^X \; I\left(\nu,p_1(E_1),p_2(E_2)\right)
\int_{z_0}^\infty N_0(z) \left| \frac{dt}{dz}\right| dz\;,
\end{equation}
where
\begin{equation}
	I(\gamma,p_1(E_1),p_2(E_2))\equiv \int_{p_1}^{p_2} \frac{p^{-\gamma+1}}{\sqrt{p^2+m_p^2}}dp\;,
\end{equation}
and where $p(E)$ is the CR proton momentum for a given kinetic energy per nucleon E.
This function can be integrated as a hyperbolic function, but for
the purpose of a simplified calculation, it is good enough to
consider the two limits, that apply either when the production
is at very low energy (e.g. for \6Li), or at GeV/n energies:
\begin{equation}
	I_{p\ll m_p} \approx \left[ \frac{p^{-\gamma+2}}{2-\gamma}\right]_{p_1}^{p_2} 
	\quad {\rm and} \quad 	
	I_{\beta \sim 1} \approx \left[ \frac{p^{-\gamma+1}}{1-\gamma}\right]_{p_1}^{p_2}\;. 
  \label{eq:Iapprox}
\end{equation}

We shall now write a simplified expression for the comparison of the production
of $X$ with respect to \6Li. The integration over $z$ cancels out,
and this results in
\begin{equation}
\left[\frac{X}{\6Li}\right] = 
   \frac{ \displaystyle\sum_{i=CR} \; \sum_{j=IGM} 
	\bar{f}_j \; F_i \;\, \sigma_{ij}^X \,\;\, I^X_{i}}{\displaystyle
	\bar{f}_{He} \; (F_{He}/4) \,\;\, \sigma_{\alpha\alpha}^{^6Li} \,\;\,
	I_{\alpha}^{^6Li}}\;,
  \label{eq:efficiency_simpl}
\end{equation}
where we wrote $I(\gamma,p_1(E^i_1),p_2(E^i_2))\equiv I^X_{i}$ for short.


\end{document}